\documentclass[8pt,twocolumn,showpacs,preprintnumbers,amsmath,amssymb,prb,superscriptaddress]{revtex4}
\usepackage{txfonts}
\usepackage{amssymb}

\usepackage{graphicx}
\usepackage{dcolumn}
\usepackage{bm}

\begin{document}

\title {How does ``Which Way" Detector affect Fano Effect in Mesoscopic Transport\\---``Phonon-Fano" Or Not}
\author{Hai-Zhou Lu}\email{luhaizhou@gmail.com}
\affiliation{Center for Advanced Study, Tsinghua University, Beijing
100084, China}
\author{Zuo-zi Chen}
\affiliation{Center for Advanced Study, Tsinghua University, Beijing
100084, China}
\author{Rong L\"{u}}
\affiliation{Center for Advanced Study, Tsinghua University, Beijing
100084, China}
\author{Bang-fen Zhu}\email{bfz@tsinghua.edu.cn}
\affiliation{Center for Advanced Study, Tsinghua University, Beijing
100084, China} \affiliation{Department of Physics, Tsinghua
University, Beijing 100084, China}

\date{\today}

\begin{abstract}
We investigate the Fano interference in the presence of a ``which
way" detector, i.e., a local electron-phonon coupler, in the context
of mesoscopic transport. Special attention is paid to study whether
the phonon sidebands in the differential conductance spectra exhibit
the typical lineshape of the Fano interference. When using a
double-dot model we obtain two seemly contradictory results by
slightly different approaches: The Markovian approach leads to {\it
absolutely no} Fano interference at the phonon sidebands, while the
Non-Markovian approach results in {\it finite} Fano interference at
the phonon sidebands. On the other hand, by using the usual
Aharonov-Bohm (AB)-ring model, i.e., only one dot is embedded into
one arm of an AB-ring, only the Non-Markovian results are recovered.
We explain these contradictory results and make a comparison between
the double-dot model and the usual AB-ring model at length.
Moreover, we also point out the essential difference between the
manifestation of the ``which way" effect in mesoscopic double-slit
experiments and its optical counterpart.

\end{abstract}

\maketitle

\section{Introduction}

Thanks to rapid advances in nanotechnology, quantum coherence can be
well maintained while electrons transport through a variety of
mesoscopic systems.~\cite{Sohn1997} Many coherent effects can thus
be observed and investigated in mesoscopic systems. For instance, in
a double-slit setup realized by using an Aharonov-Bohm (AB) ring
with a quantum dot (QD) embedded,~\cite{Yacoby1995,Schuster1997} the
interference between the resonant channel (the QD) and the continuum
channel (the reference arm) will result in a typical asymmetric Fano
resonance in the differential conductance
spectra,~\cite{Kobayashi2002} because the phase of an electron would
change by $\pi$ when crossing a resonant level, compared with little
phase shift in the continuum channel.

\begin{figure}[htbp]
\centering
\includegraphics[width=0.4\textwidth]{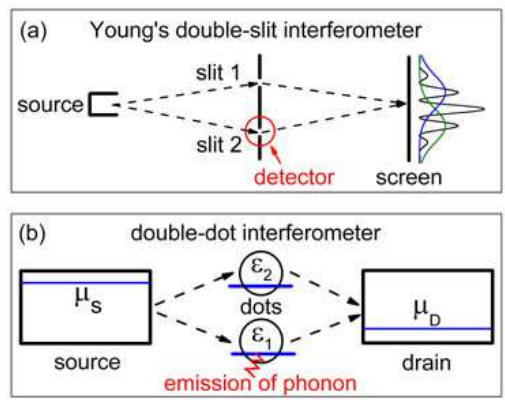}
\caption{(Color online)(a) Illustration of the ``which way" effect
in the Young's double-slit experiment.
(b) Configuration of the double-dot interferometer, where only dot 1
is coupled to a local optical phonon mode.}\label{fig:configure}
\end{figure}

As a vivid illustration of the Bohr's complementarity principle in
quantum mechanics, in an optical Young's double-slit experiment as
shown in Fig.~\ref{fig:configure}(a) if one detects which slit the
photon passes through (even in principle), the interference pattern
on the manifest screen will be destroyed. Candidates to realize such
``which way" detector in mesoscopic systems in principle can be a
diversity of scattering mechanisms, such as the electron-electron
interaction induced spin flip~\cite{Aikawa2004} and Coulomb
repulsion,~\cite{Buks1998} the light radiation,~\cite{Jauho1998} the
electron-phonon interaction (EPI),~\cite{Park2002,Yu20043,LeRoy2004}
{\it etc.} Since these mechanisms play important roles in transport
in mesoscopic systems, to study how the ``which way" detector
affects the interference (e.g. Fano effect) would be crucial to
understand the fundamental decoherence process in mesoscopic systems
and would shed light on the potential applications.

To address this subject, we introduce a double-dot
interferometer~\cite{Kubala2002,Marquardt2003} as depicted in
Fig.~\ref{fig:configure}(b). Specifically, the QD here is assumed to
be a single-molecule coupled to two metallic electrodes. The phonon
sidebands in the differential conductance are introduced by strong
coupling between the electron and local optical vibration mode
within the
single-molecule.~\cite{Reed1997,Nitzan2000,Stipe1998,Park2000,Zhitenev2002,Yu20043,Pasupathy2005,Sapmaz2006}
This model is ideal to deal with the ``which way" problem: The
theoretical derivation is much simplified compared to the usual
one-dot AB-ring setup adopted in most
works~\cite{Entin-Wohlman2004,Ueda2006} yet captures the essential
physics. For general purpose, in the end we will compare this
double-dot model with the usual AB-ring model in which only one QD
is embedded in the AB-ring arms. We will see that, with minor
modifications, our results can apply equally well to other
interference models with the setup changed or the ``which way"
detector replaced.

As far as we know, most of previous publications mainly discuss how
the ``which way" detector affects the interference between elastic
channels (the zero-phonon channel in our case), such as the
visibility of AB oscillation\cite{Marquardt2003} or the lineshape of
the Fano resonance,~\cite{Ueda2006}{\it etc}; few attentions have
been paid to the coherence of inelastic channels (phonon-assisted
channels here).~\cite{Entin-Wohlman2004,Ma2004,Ueda2006} Regarding
this, in this article we intend to investigate the coherence through
the inelastic channels by examining whether the Fano interference
survives in the differential conductance spectra after a phonon is
emitted (absorbed). The concern over the Fano lineshape in phonon
sidebands has already been studied in Raman spectra of doped
semiconductors.~\cite{Jauho1984,Ager1994} We shall show that by two
slightly different approaches two incompatible results are obtained
in the conductance spectra of the double-dot model. By the first
approach that deals with the Markovian case, the phonon sidebands
give {\it absolutely no} hint of the Fano interference; while by the
second approach that suits to the Non-Markovian case, the phonon
sidebands do exhibit Fano interference to certain extent. On the
other hand, for the one-dot AB-ring setup, only the Non-Markovian
results can be recovered.~\cite{Entin-Wohlman2004, Ueda2006}
Compared to the usual Young's double-slit experiment, it seems that
the Markovian result agrees with the ``which way" rule, while the
Non-Markovian result violates it. By carefully examining the
electron-phonon decoupling process, we have found the key to these
incompatible results. We will show that the usual AB-ring model only
works for one circumstance while the double-dot model can capture
both. We will also point out that the coherent multi-reflection
would be responsible for the different manifestations of ``which
way" effect in mesoscopic double-slit interferometer and its optical
analogy.

\section{Model} The Hamiltonian for the double-dot interferometer
depicted in Fig.~\ref{fig:configure}(b) can be expressed as
\begin{eqnarray}
{\bf H}=\sum_{{\bf k},\eta\in\text{S,D}}\varepsilon_{\eta{\bf
k}}{\bf c}^{\dag}_{\eta{\bf k}}{\bf c}_{\eta{\bf
k}}+\sum_{i\in1,2}\varepsilon_i{\bf d}^{\dag}_i{\bf d}_i+\lambda{\bf
d}^{\dag}_1{\bf d}_1({\bf a}^{\dag}+{\bf a})\nonumber\\
+\hbar\omega_0{\bf a}^{\dag}{\bf a}+\sum_{\eta{\bf
k}i}\left(V_{\eta{\bf k}i}{\bf c}^{\dag}_{\eta{\bf k}}{\bf
d}_i+h.c.\right).\label{Equation:H}
\end{eqnarray}
Here  ${\bf c}^{\dag}_{\eta{\bf k}}$ (${\bf c}_{\eta{\bf k}}$) is
the creation (annihilation) operator for an electron with wave
vector {\bf k} in the lead $\eta$ [ $\eta=\mathrm{S(D)}$ stands for
the source (drain) lead], and ${\bf d}^{\dag}_i$ (${\bf d}_i$) is
the creation (annihilation) operator of an electron residing in the
dot $i$. The spin index is omitted here for simplicity. Phonon
operator ${\bf a}^{\dag}$ (${\bf a}$) creates (annihilates) a local
optical vibration in dot $1$ at frequency $\omega_0$ with the EPI
strength $\lambda$. Electron can tunnel between the lead $\eta$ and
dot $i$ with a hopping amplitude denoted as $V_{\eta{\bf k}i}$. By
denoting the chemical potential in each lead as $\mu_{\eta}$, we
express the gate voltage as
$V_{\text{mid}}=(\mu_{\text{S}}+\mu_{\text{D}})/2e$, and the bias
voltage as $V_{\text{bias}}=(\mu_{\text{S}}-\mu_{\text{D}})/e$.

Within the nonequilibrium Green function formalism the steady
current tunneling through the interferometer can be expressed
as,~\cite{Jauho1994}
\begin{eqnarray}
J=J^{\text{sum}}+J^{\text{inte}},
\end{eqnarray}
where
\begin{eqnarray}
J^{\delta}\equiv(J^{\delta}_{\text{S}}-J^{\delta}_{\text{D}})/2,
\label{Equation:J}
\end{eqnarray}
with $\delta$=\text{sum},\text{inte}. For later convenience of
studying the interference phenomena, we separate the total current
$J$ into two parts: the sum of currents through two channels
$J^{\text{sum}}$, and the interference between two channels
$J^{\text{inte}}$, i.e.,
\begin{eqnarray}
J_{\eta}^{\text{sum}}=\frac{e}{h}\int
d\omega\sum_{i}\Gamma^{\eta}_{ii}(\omega)
\{f_{\eta}(\omega)A_{ii}(\omega)
+iG^<_{ii}(\omega)\},\label{Equation:Jsum}\\
J_{\eta}^{\text{inte}}=\frac{e}{h}\int
d\omega\sum_{i}\Gamma^{\eta}_{i\bar{i}}(\omega)
[f_{\eta}(\omega)A_{\bar{i}i}(\omega)
+iG^<_{\bar{i}i}(\omega)],\label{Equation:Jinter}
\end{eqnarray}
where $G^{r(a)}_{ij}$ and $G^{>(<)}_{ij}$ are respectively the
retarded (advanced) and the greater (lesser) Green functions of the
dot region, the index $\bar{i}=2(1)$ if $i=1(2)$, $A_{ij}(\omega)$
is the spectral function defined via $A_{ij}=i(G^>_{ij}-G^<_{ij})$,
and $\Gamma^{\eta}_{ij}(\omega)$ is the broadening function defined
as $2\pi\sum_{\bf k}V_{\eta{\bf k} i}^*V_{\eta{\bf k}
j}\delta(\varepsilon_{\eta{\bf k}}-\omega)$. In the following, we
shall derive these Green functions by using the equation of motion
method and disentangle the electron-phonon part via the canonical
transformation. The differential conductance, $\mathbf{G}\equiv
\mathbf{G}^{\text{sum}}+\mathbf{G}^{\text{inte}}$, can be obtained
accordingly via
\begin{eqnarray}
\mathbf{G}^{\text{sum}}=\partial_{V_\text{bias}}J^{\text{sum}},\qquad
\mathbf{G}^{\text{inte}}=\partial_{V_\text{bias}}J^{\text{inte}}.\label{Equation:Con}
\end{eqnarray}
We shall show that each peak in the summation term
$\mathbf{G}^{\text{sum}}$ corresponds to a channel that electron
tunnels through, and the hump and dip structure at corresponding
energy in the interference term $\mathbf{G}^{\text{inte}}$ can be
used to identify whether two channels interfere with each other or
not.

\section{Fano Effect Without EPI ($\lambda=0$)} In the absence
of EPI, we can reduce the Hamiltonian in Eq.~(\ref{Equation:H}) to
the Fano-Anderson Hamiltonian, whereby we obtain the retarded
Green's functions rigorously by the equation of motion approach,
\begin{eqnarray}
G^{0,r}_{ij}(\omega)=\frac{1} {Q}\left(\begin{array}{cc}
\omega-\varepsilon_2+i\Gamma_{22}/2,&-i\Gamma_{12}/2\\
-i\Gamma_{21}/2,&\omega-\varepsilon_1+i\Gamma_{11}/2
\end{array}\right),\label{Equation:G0}
\end{eqnarray}
where
$$\Gamma_{ij}\equiv\Gamma^{\text{S}}_{ij}+\Gamma^{\text{D}}_{ij},$$
and
$$Q\equiv(\omega-\varepsilon_1+i\Gamma_{11}/2)(\omega-\varepsilon_2+i\Gamma_{22}/2)+\Gamma_{12}\Gamma_{21}/4.$$
The superscript zero of the Green's function here denotes the
absence of EPI ($\lambda=0$). We then obtain the spectral function
$A^{0}_{ij}(\omega)=-2\Im\{G^{0,r}_{ij}(\omega)\}$, the advanced
Green's functions $G^{0,a}_{ji}(\omega)=(G^{0,r}_{ij}(\omega))^*$,
and the greater (lesser) Green's functions via the Keldysh formula
as
$G^{0,>(<)}_{ij}(\omega)=\sum_{mn}G^{0,r}_{im}(\omega)\Sigma^{0,>(<)}_{mn}(\omega)G^{0,a}_{nj}(\omega)$,
where
$\Sigma^{0,<}_{ij}(\omega)=i\sum_{\eta}f_{\eta}(\omega)\Gamma^{\eta}_{ij}$,
and $\Sigma^{0,>}_{ij}=\Sigma^{0,<}_{ij}-i\Gamma_{ij}$. By
substituting the expressions above into
Eqs.(\ref{Equation:J})-(\ref{Equation:Con}) we obtain the required
currents and conductances.

\begin{figure}[htbp]
\centering
\includegraphics[width=0.48\textwidth]{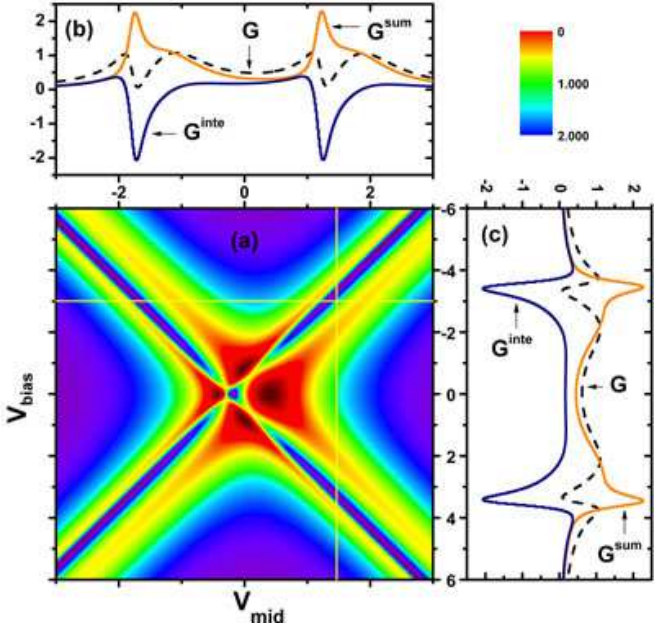}
\caption{(Color online)(a) The zero-temperature differential
conductance $\mathbf{G}$ as a function of $V_{\text{bias}}$ and
$V_{\text{mid}}$ in the absence of EPI. (b) and (c) are section
profiles for fixed $V_{\text{bias}}$ and $V_{\text{mid}}$,
respectively. Here the interference component $\mathbf{G}^{inte}$
(black solid lines) of the total conductance $\mathbf{G}$ (dashed
lines) as well as the sum of the conductance through dot 1 and 2
($\mathrm{G^{\mathrm{sum}}}$, yellow solid lines) are plotted in the
same panels for comparison. Throughout the paper the phonon
frequency $\hbar\omega_0$ is set as the energy unit. Other
parameters used are $\varepsilon_1=-0.4$, $\varepsilon_2=0.4$,
$\Gamma_{11}=0.2$, $\Gamma_{22}=0.6$,
$\Gamma_{12}=\Gamma_{21}=\sqrt{\Gamma_{11}\Gamma_{22}}$.}\label{fig:nonepi}
\end{figure}

The zero-temperature differential conductance through the double
dots without EPI is plotted in Fig.~\ref{fig:nonepi} as a function
of the bias voltage $V_{\text{bias}}$ and gate voltage
$V_{\text{mid}}$. The typical asymmetric lineshape in the
differential conductance is the manifestation of the Fano
interference between two channels through the double dots, which has
already been extensively discussed in the previous
publication.~\cite{Kubala2002,Guevara2003,Lu2005} The interference
pattern will switch from the constructive to the destructive
whenever the incident electron experiences a $\pi$ phase change in
channel 1 ( through crossing the resonance level $\varepsilon_1$ in
dot 1) while remains its phase in the reference channel (dot 2). We
also plot the corresponding $\mathbf{G}^{\text{sum}}$ and
$\mathbf{G}^{\text{inte}}$ in the section profiles for comparison.
As clearly shown by the curves, every time $\mathbf{G}$ exhibits the
asymmetric Fano lineshape,  the interference term
$\mathbf{G}^{\text{inte}}$ will exhibit the hump and dip structure,
which can also be viewed as an evidence of the Fano interference.

\section{Fano Effect With EPI ($\lambda\neq0$)}

In order to treat the EPI unperturbatively, we employ the standard
canonical transformation technique~\cite{Mahan1990} to eliminate the
electron-phonon coupling terms in the Hamiltonian
Eq.~(\ref{Equation:H}). Namely, let ${\bf S}\equiv\lambda{\bf
d}_1^{\dag}{\bf d}_1\left({\bf a}^{\dag}-{\bf
a}\right)/\hbar\omega_0$, the transformed Hamiltonian reads
$$\bar{\bf H}= e^{\bf S}{\bf H}e^{-{\bf S}} \equiv \bar{\bf
H}_{\text{ph}}+\bar{\bf H}_{\text{el}}.$$ In the transformation, the
phonon part keeps unchanged, while the electron part is reshaped
into
\begin{eqnarray}
\bar{\bf H}_{\text{el}}=\sum_{\eta{\bf k}}\varepsilon_{\eta{\bf
k}}{\bf c}^{\dag}_{\eta{\bf k}}{\bf c}_{\eta{\bf
k}}+\sum_i\bar{\varepsilon}_i{\bf d}^{\dag}_i{\bf d}_i\
+\sum_{\eta{\bf k}i}\left(\bar{V}^{\eta}_{{\bf k}i}{\bf
c}^{\dag}_{\eta{\bf k}}{\bf d}_i+h.c.\right),
\end{eqnarray}
where $\bar{\varepsilon}_2 = \varepsilon_2$,
$\bar{\varepsilon}_1\equiv \varepsilon_1-\lambda^2/\hbar\omega_0$,
 $\bar{V}^{\eta}_{{\bf
k}2} = V^{\eta}_{{\bf k}2}$, and $\bar{V}^{\eta}_{{\bf k}1}\equiv
V^{\eta}_{{\bf k}1}{\bf X}\equiv V^{\eta}_{{\bf k}1}
\exp\left(-\lambda({\bf a}^{\dag}-{\bf a})/\hbar\omega_0\right)$.
It is noticed that the transformed hopping amplitude contains the
phonon operator ${\bf X}$, indicating that the electron and phonon
are still entangled during hopping processes. The usual practice as
a first-order approximation is replacing the phonon operator with
its mean-field expectation value, i.e.,
\begin{eqnarray}
\bar{V}_{{\bf k}\eta,1}\approx V_{{\bf k}\eta,1}\langle{\bf
X}\rangle=V_{{\bf k}\eta,1}e^{-\phi_0},\label{Equation:MeanField}
\end{eqnarray}
where $\phi_0\equiv g(N_{\text{ph}}+1/2)$,
$g\equiv(\lambda/\omega_0)^2$ and $N_{\text{ph}}$ denotes the phonon
occupation which obeys the Bose distribution at temperature $T$. By
this polaron approximation, electrons are decoupled from phonons,
and $\bar{\bf H}_{\text{el}}$ is transformed into the Fano-Anderson
Hamiltonian with the parameters renormalized. Hereafter we denote
the Green's functions defined in the context of $\bar{\bf
H}_{\text{el}}$ by adding a small bar overhead, i.e.,
$\bar{G}_{ij}$, which can be readily expressed with $G^0_{ij}$ in
Eq.~(\ref{Equation:G0}) by replacing the superscript zero with a
small bar.

In the following, we shall discuss two slightly different approaches
to evaluate the full Green functions $G^{r(a,>,<)}_{ij}$ from the
$\bar{G}^{r(a,>,<)}_{ij}$.

\begin{figure}[htbp]
\centering
\includegraphics[width=0.48\textwidth]{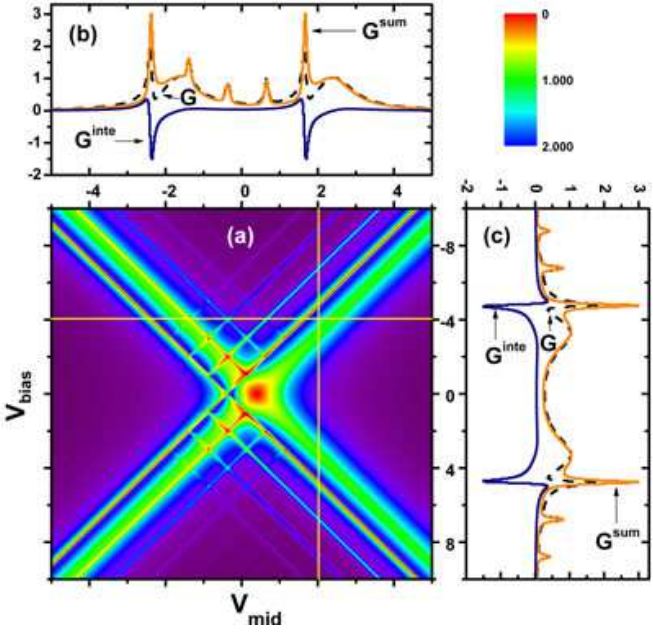}
\caption{(Color online) The differential conductance obtained by the
Markovian approach in the presence of EPI ($\lambda=1$), with the
parameters and line notation same with
Fig.~\ref{fig:nonepi}.}\label{fig:first}
\end{figure}

The first one is straightforward.~\cite{Lu2006} By using
Eq.~(\ref{Equation:MeanField}) and the relation like $\langle{\bf
d}_i(t){\bf d}^{\dag}_j(t')\rangle=\langle\bar{\bf d}_i(t)\bar{\bf
d}^{\dag}_j(t')\rangle \langle
\mathbf{X}(t)\mathbf{X}^{\dag}(t')\rangle$, we decouple the Green
functions\cite{Chen2005} and obtain the intra-dot Green functions as
\begin{eqnarray}
G_{11}^{>(<)}(\omega)&=&\sum_nL_n\bar{G}_{11}^{>(<)}(\omega\mp
n\omega_0),\nonumber\\
G_{22}^{>(<)}(\omega)&=&\bar{G}_{22}^{>(<)}(\omega).\label{Equation:G><dia}
\end{eqnarray}
Here
$$L_n\equiv e^{n\omega_0 \beta/2}e^{-(g(2N_{\text{ph}}+1)}I_n
\left[2g\sqrt{N_{\text{ph}}(N_{\text{ph}}+1)}\right],$$ in which
 $I_n(z)$ is the $n$-th Bessel function
of complex argument $z$, and $\beta=1/k_{\text{B}}T$. Since only the
electron in dot 1 is subjected to the EPI, $\langle{\bf d}_1(t){\bf
d}^{\dag}_2(t')\rangle=\langle\bar{\bf d}_1(t){\bf
d}^{\dag}_2(t')\rangle \langle \mathbf{X}\rangle$, and the inter-dot
Green functions then read
\begin{eqnarray}
G_{12}^{>(<)}=\bar{G}^{>(<)}_{12}e^{-\phi_0},\qquad
G_{21}^{>(<)}=\bar{G}^{>(<)}_{21}e^{-\phi_0}.\label{Equation:G><off1}
\end{eqnarray}
The differential conductances with finite EPI obtained by the first
approach are plotted in Fig.~\ref{fig:first}. Compared with the
non-EPI case, both $\mathbf G$ and $\mathbf G^{\text{sum}}$
demonstrate a set of sidebands at $\bar{\varepsilon}_1\pm
n\hbar\omega_0$, which are mathematically related with the phonon
expansions in $G^{>(<)}_{11}(\omega)$ and come from the phonon
correlation functions $\langle{\bf X}(t){\bf
X}^{\dag}(t')\rangle_{\text{ph}}$ or the $\langle{\bf
X}^{\dag}(t'){\bf X}(t)\rangle_{\text{ph}}$. However, the inter-dot
correlation functions obtained in this way only contain $\langle{\bf
X}(t)\rangle_{\text{ph}}$ or $\langle{\bf
X}^{\dag}(t')\rangle_{\text{ph}}$, thus
$G^{>(<)}_{i\bar{i}}(\omega)$ can not be expanded into similar
phonon series. As a result, besides the hump and dip near
$\bar{\varepsilon}_1$, $\mathbf{G}^{\text{inte}}$ exhibits no
side-peak structure at $\bar{\varepsilon}_1\pm n\hbar\omega_0$, and
$\mathbf{G}$ shows no Fano interference feature at the phonon
sidebands either. This means that the phonon channel through dot 1
does not interfere with the reference channel through dot 2, unlike
the zero-phonon channel.

If we assume that no phonon is involved in the tunneling process of
electron through dot 2, the result above seems reasonable and agrees
with the ``which way" rule. In this case, phonon-assisted channel is
in principle recognized if the emitted or absorbed phonon is
detected. As a result the phonon-assisted channel will not interfere
with the elastic reference channel, and $G$ shows no interfere
feature at phonon side-peaks. However, these statements are true
only under the assumption that the electron is {\it not} coupled to
phonons {\it at all} when passing the reference channel, as in the
case of Young's two-slit interfere experiment. To justify whether
the assumption is valid, we need to clarify two different situations
in usual mesoscopic systems, which critically depends on the
coherent properties of the leads: (i) The Markovian leads, into
which the electrons will lose their phase memory completely, namely
the electrons which originally come from dot 1 forget their previous
coupling with phonons once entering into both leads and dot 2; (ii)
The Non-Markovian leads, into which the electron still keep its
phase memory to some extent, namely the electron in dot 2 will be
coupled to phonons as the electron in dot 1 through a coherent
multi-reflection process. Clearly the first approach is the
Markovian approach, which suits for the cases with the Markovian
leads and certainly not for the Non-Markovian leads.

\begin{figure}[htbp]
\centering
\includegraphics[width=0.48\textwidth]{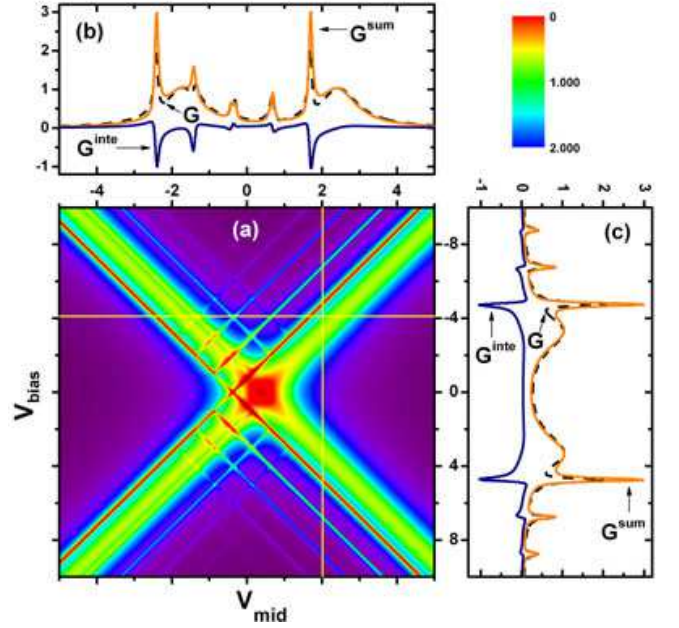}
\caption{(Color online) The differential conductance obtained by the
Non-Markovian approach in the presence of EPI ($\lambda=1$), with
the same parameters and line notation as in
Fig.~\ref{fig:first}.}\label{fig:Second1}
\end{figure}

To account for the Non-Markovian lead, we modify the first approach
a little bit. More specific, we will modify the way of decoupling
the electrons from phonons. The previous usage of
Eq.~(\ref{Equation:MeanField}) decouples the lead electrons from
phonons completely, thus no entanglement between the electron in dot
2 and phonons needs considering in subsequent derivations. To avoid
this for the Non-Markovian case, we shall derive the inter-dot
correlation function before taking the electron-phonon decoupling
approximation. In the following, we shall outline our derivation for
the Non-Markovian case, but leave the details in the Appendix.

Firstly, with the help of the equation of motion we can express the
inter-dot correlation functions $G^{>(<)}_{ij}$ in terms of the
intra-dot Green function $G^{>(<)}_{ii}$ as
\begin{eqnarray}
G^{>(<)}_{12}(\omega)&=&G^{>(<)}_{11}(\omega)s^a_{12}(\omega)+G^r_{11}(\omega)p^{>(<)}(\omega),\nonumber\\
G^{>(<)}_{21}(\omega)&=&s^r_{21}(\omega)G^{>(<)}_{11}(\omega)+q^{>(<)}(\omega)G^a_{11}(\omega),\label{Equation:G>21}
\end{eqnarray}
where the functions $s^{r(a)}_{ij}$ and $q(p)^{>(<)}$ can be found
in the Appendix \ref{Appendix:interdot}. Secondly, by using the same
electron-phonon decoupling approximation
[Eq.~(\ref{Equation:MeanField})] and similar procedure as in the
Markovian approach, we get the intra-dot Green functions
$G^{>(<)}_{ii}$ as in Eq.~(\ref{Equation:G><dia}) and
$G^{r(a)}_{11}$ via
\begin{eqnarray}
G^{r(a)}_{ii}(\omega)=\textsl{P}\int_{-\infty}^{\infty}\frac{d\omega'}{2\pi}\frac{A_{ii}(\omega)}{\omega-\omega'}\mp
\frac{i}{2}A_{ii}(\omega).\label{Equation:Grdia}
\end{eqnarray}
Finally, we substitute $G^{r(a,>,<)}_{11}$ back into
Eq.~(\ref{Equation:G>21}) and obtain all the required Green
functions. It is noticed that by this modified decoupling procedure,
the inter-dot correlation functions also contain the terms of
$\langle{\bf X}^{\dag}(t){\bf X}(t')\rangle_{\text{ph}}$ and can be
expanded into phonon series. Therefore the entanglement between
phonons and electrons in dot 2 are partially maintained via coherent
multi-reflection.

Differential conductance calculated with the Non-Markovian approach
is plotted in Fig.~\ref{fig:Second1}. Compared with the Markovian
results in Fig.~\ref{fig:first}, we have found $\mathbf
G^{\text{sum}}$ remains unchanged, while $\mathbf G^{\text{inte}}$
contains some phonon structures that are totally absent in
Fig.~\ref{fig:first}. This indicates that the phonon-assisted
tunneling channel through dot 1 does interfere with the reference
channel through dot 2, so certain Fano lineshape appears in the
phonon side-peaks of the differential conductance $\mathbf G$, as
obtained in previous results in the AB-ring
setup.~\cite{Entin-Wohlman2004,Ueda2006}

\begin{figure}[htbp]
\centering
\includegraphics[width=0.4\textwidth]{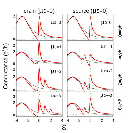}
\includegraphics[width=0.4\textwidth]{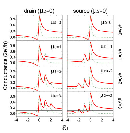}
\caption{(Color online) Comparison between results of the double-dot
interferometer by the non-Markovian approach (upper 8 panels) and
the AB-ring model (lower 8 panels). The conductances contributed
from the source and drain leads are plotted in the left and right
panels, respectively, in which the thick solid lines represent the
total conductance, the thin dash lines stand for the conductance
when only the dot 1 channel is open, and the thin solid lines are
the conductance when only the dot 2 in the double-dot system or the
reference channel in the AB-ring model is conducting. The parameters
for the double-dot are
$3\Gamma_{11}^{\mathrm{S/D}}=\Gamma_{22}^{\mathrm{S/D}}=0.5$,
$\varepsilon_2=\overline{\varepsilon}_1+3$, and that for the AB-ring
are $\Gamma_{\mathrm{S}}=\Gamma_{\mathrm{D}}=0.5$, $\xi=0.05$,
$\phi=0$. }\label{fig:compare 0}
\end{figure}

\begin{figure}[htbp]
\centering
\includegraphics[width=0.4\textwidth]{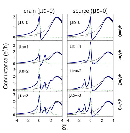}
\includegraphics[width=0.4\textwidth]{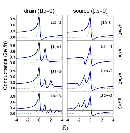}
\caption{(Color online) The same as in Fig.~\ref{fig:compare 0}
except $\phi=\pi$ for the AB-ring model and
$\varepsilon_2=\overline{\varepsilon}_1-3$ for the double-dot model.
}\label{fig:compare pi}
\end{figure}

\section{Comparison between AB-ring and double-dot with Non-Markovian approach}
Above results are generally true for other multi-channel interfere
systems with local ``which way" detector. For example, we calculate
the differential conductance in a usual AB-ring model as given in
Appendix \ref{Appendix:ABring}, and compare the results with the
double-dot interference system calculated with the Non-Markovian
approach (Figs. \ref{fig:compare 0} and \ref{fig:compare pi}). For
better comparison to the previous work,~\cite{Ueda2006} we plot the
differential conductance spectra as functions of
$(\overline{\varepsilon}_{1}-\mu_{\mathrm{S/D}})$ at several fixed
bias voltage $eV=\mu_{\mathrm{S}}-\mu_{\mathrm{D}}$, which is
increased from the top to the bottom in the figures. It is worth
pointing out that here we have used the improved electron-phonon
decoupling scheme for the Green functions,~\cite{Chen2005} so we can
deal with the bias-induced phonon-assisted processes at low
temperature more appropriately. As a consequence, the differential
conductance at zero temperature exhibit phonon sidepeaks at both
sides of the zero-phonon peaks, which is quite different from the
previous work.~\cite{Ueda2006} To clearly show the improvements, we
measure the conductance by the source and drain Fermi levels on the
left and right columns, respectively. Technically, they are obtained
by $\lim_{\delta\rightarrow 0}[I(\overline{\varepsilon}_{1},
\mu_{\mathrm{D}})-I(\overline{\varepsilon}_{1},
\mu_{\mathrm{D}}-\delta)]/\delta $ and $\lim_{\delta\rightarrow
0}[I(\overline{\varepsilon}_{1}, \mu_{\mathrm{S}}+\delta
)-I(\overline{\varepsilon}_{1}, \mu_{\mathrm{S}})]/\delta $.

As shown in Figs. \ref{fig:compare 0} and \ref{fig:compare pi},  in
the AB-ring model the Fano-asymmetric lineshapes also appear in the
phonon sidepeaks of the differential conductance, just like the
Non-Markovian case for the double-dot model. The reason why the
AB-ring model's result is similar to the Non-Markovian case rather
than the Markovian case lies in its current formula
Eq.~(\ref{transmission}), which is obtained by integrating out the
reference channels to get a formula in terms of the dot Green
functions before taking any approximation to EPI. This procedure is
essentially equivalent to the Non-Markovian approach we have adopted
in the double-dot model, so it is no wonder why the AB-ring's result
exhibits the similar Phonon-Fano lineshapes. Moreover, since the
reference channel in the AB-ring model connects two leads directly,
it is not easy to take into account the Markovian effect in the
AB-ring model in contrasts to the double-dot model.

The main difference between two models is that the transmission
probability through the reference channel is constant in the AB-ring
model, while it varies as a function of energy in the double-dot
model. Despite of this difference, one can easily identify the
similarities between the results of
$\varepsilon_2=\overline{\varepsilon}_1+3$ (double-dot model) and
$\phi=0$ (AB-ring model) in Fig.~\ref{fig:compare 0}, or
$\varepsilon_2=\overline{\varepsilon}_1-3$ (double-dot model) and
$\phi=\pi$ (AB-ring model) in Fig.~\ref{fig:compare pi}, because due
to the Friedel's sum rule~\cite{Friedel1953} a $\pi$ phase shift
occurs for electron in dot 1 when its incident energy crosses
$\overline{\varepsilon}_1$, which is equivalent to adding a $\pi$
magnetic flux threading the AB-ring.

\section{Conclusion} In summary, the Fano effect in the presence
of local electron-phonon coupling has been studied in a double
quantum dot system compared to an AB-ring system. In contrast to its
optical analogy, when tunneling into the leads in a mesoscopic
systems, electrons can either lose their phase memory completely
(Markov case) or maintain at least part of the coherence (Non-Markov
case), which determines whether the coherent multi-reflection
between different channels exists. As a consequence, the phonon
sidebands in the differential conductance will not show the Fano
interference for the Markov cases, or exhibit the Fano interference
to certain extent for the Non-Markov cases. The double-dot model can
describe both circumstances while the AB-ring model can only capture
the Non-Markovian case. In the double-dot model, to account for both
cases, the usual electron-phonon decoupling approximation should be
treated with cares.

{\it Acknowledge:} The authors would like to acknowledge Dr. Hui
Zhai, Chao-Xing Liu, Prof. M. Eto and Dr. A. Ueda for helpful
discussions. This research is supported by the NSFC (Grant
10574076), the Basic Research Program of China (Grant No.
2006CB921500, 2006CB605105) and the MOE of China (Grant No.200221).

\appendix

\section{Derivation of the inter-dot correlation
functions}\label{Appendix:interdot}

The inter-dot correlation functions can be derived as functions of
the intra-dot Green functions by using the equations of motion
method,
\begin{eqnarray}
\left(-i\partial_{\tau'}-\varepsilon_2\right)G_{12}(\tau,\tau')=\sum_{\eta{\bf
k}}V_{\eta{\bf k}2}G_{1,\eta{\bf k}}(\tau,\tau').
\end{eqnarray}
The factor $\left(-i\partial_{\tau'}-\varepsilon_2\right)$ can be
interpreted as the inverse operator of the noninteracting electron
Green function of dot 2 multiplied from left, then we have
\begin{eqnarray}
G_{12}(\tau,\tau')=\sum_{\eta{\bf k}}V_{\eta{\bf k}2}\int
d\tau_1G_{1,\eta{\bf k}}(\tau,\tau_1)g_2(\tau_1,\tau').
\end{eqnarray}
Similarly,
\begin{eqnarray}
G_{1,\eta{\bf k}}(\tau,\tau_1)=\sum_jV^*_{\eta{\bf k}j}\int
d\tau_2G_{1j}(\tau,\tau_2)g_{\eta{\bf k}}(\tau_2,\tau_1).
\end{eqnarray}
Combine these two equations, we have
\begin{eqnarray}
G_{12}(\tau,\tau')=\sum_{j}\int d\tau_1 d\tau_2
G_{1j}(\tau,\tau_2)\Sigma_{j2}(\tau_2,\tau_1)g_2(\tau_1,\tau'),\nonumber
\end{eqnarray}
where the self energy is defined as
\begin{eqnarray}
\Sigma_{j2}(\tau_2,\tau_1)\equiv\sum_{\eta{\bf k}}V^*_{\eta{\bf
k}j}g_{\eta{\bf k}}(\tau_2,\tau_1)V_{\eta{\bf k}2}.
\end{eqnarray}
To do analytic continuation for terms like $D=\int_C ABC$, we have
the Langreth Theorem, i.e.,
\begin{eqnarray}
D^{>}&=&\int_t\left(A^rB^rC^{>}+A^rB^{>}C^a+A^{>}B^aC^a\right),\nonumber\\
D^{r}&=&\int_t A^{r}B^{r}C^{r}.
\end{eqnarray}
Thus, after the analytic continuation and Fourier transformation, we
have
\begin{eqnarray}
G^<_{12}(\omega)&=&\sum_jG^r_{1j}(\omega)\left[\Sigma^r_{j2}(\omega)g^<_2(\omega)+\Sigma^<_{j2}(\omega)g^a_2(\omega)\right]\nonumber\\
&&\ \ \ \ \ \ \ \ \ \ \ \ \ \ \ \ \ \ \ \ \ \ \ \ \ \ \ \ \  +\sum_jG^<_{1j}\Sigma^a_{j2}(\omega)g^a_2(\omega),\nonumber\\
G^r_{12}(\omega)&=&\sum_jG^r_{1j}(\omega)\Sigma_{j2}^r(\omega)g_2^r(\omega).
\end{eqnarray}
Further simplification shows that
\begin{eqnarray}
G^<_{12}(\omega)&=&G^<_{11}(\omega)s^a_{12}(\omega)+G^r_{11}(\omega)p^<(\omega),\\
G^{r}_{12}(\omega)&=&G^{r}_{11}(\omega)s^r_{12}(\omega).
\end{eqnarray}
Similarly,
\begin{eqnarray}
G^<_{21}(\omega)&=&s^r_{21}(\omega)G^<_{11}(\omega)+q^<(\omega)G^a_{11}(\omega),\\
G^a_{21}(\omega)&=&s^a_{21}(\omega)G^a_{11}(\omega).
\end{eqnarray}
After having the expressions for lesser Green functions, the greater
Green functions can then be readily obtained just by replacing the
$<$ to $>$ in the above expressions, i.e.,
\begin{eqnarray}
G^>_{12}(\omega)&=&G^>_{11}(\omega)s^a_{12}(\omega)+G^r_{11}(\omega)p^>(\omega),\\
G^>_{21}(\omega)&=&s^r_{21}(\omega)G^a_{11}(\omega)+q^>(\omega)G^r_{11}(\omega).
\end{eqnarray}
Here the correlation functions $p^{>(<)}$, $q^{>(<)}$ and
$s^{r(a)}_{ij}$ are defined as below,
\begin{eqnarray}
p^{>(<)}&\equiv&\frac{\Sigma^r_{12}g^{>(<)}_2+\Sigma^{>(<)}_{12}g^a_2}
{\left(1-\Sigma^r_{22}g^r_2\right)\left(1-\Sigma_{22}^ag^a_2\right)}+\frac{\Sigma^r_{12}\Sigma^{>(<)}_{22}-\Sigma^r_{22}\Sigma^{>(<)}_{12}}
{\left(1-\Sigma^r_{22}g^r_2\right)\left(1-\Sigma_{22}^ag^a_2\right)}g^r_2g^a_2,\nonumber\\
q^{>(<)}&\equiv&\frac{g^r_2\Sigma^<_{21}+g^{>(<)}_2\Sigma^a_{21}}{(1-g^r_2\Sigma^r_{22})(1-g^a_2\Sigma^a_{22})}
+\frac{\Sigma^a_{21}\Sigma^{>(<)}_{22}-\Sigma^a_{22}\Sigma^{>(<)}_{21}}{(1-g^r_2\Sigma^r_{22})(1-g^a_2\Sigma^a_{22})}g^r_2g^a_2,\nonumber\\
s^{r(a)}_{ij}&\equiv&\frac{\Sigma^{r(a)}_{ij}g^{r(a)}_2}{1-\Sigma^{r(a)}_{22}g^{r(a)}_2}.
\end{eqnarray}
For cases where $V_{\eta{\bf k}i}$ is independent of ${\bf k}$,
$\Sigma^{r(a)}_{ij}\Sigma^{>(<)}_{22}-\Sigma^{r(a)}_{22}\Sigma^{>(<)}_{ij}=0$
and the second term in the expressions of $p$ or $q$ vanishes.

\section{Model and Derivation of AB-ring
Model}\label{Appendix:ABring}

The Hamiltonian of the AB-ring model
reads\cite{Hofstetter2001,Ueda2006}
\begin{eqnarray}
 \mathbf{H}_{\mathrm{AB}}= \mathbf{H}_{\mathrm{leads}}+ \mathbf{H}_{\mathrm{D}}
+ \mathbf{H}_{\mathrm{T}},
\end{eqnarray}
where the lead part $\mathbf{H}_{\mathrm{leads}}$ is the same as in
the Eq.~(\ref{Equation:H}). The second term describes the single
level coupled to the single vibrational mode
\begin{eqnarray}
\mathbf{H}_{\mathrm{D}}=\varepsilon_1 \mathbf{d}^{\dag}_1
\mathbf{d}_1+\hbar\omega_0\mathbf{a}^{\dag}\mathbf{a}+\lambda
\mathbf{d}^{\dag}_1\mathbf{d}_1(\mathbf{a}^{\dag}+\mathbf{a}),
\end{eqnarray}
where the operators and parameters are of the same meaning as those
in the double-dot interferometer. The last term
\begin{eqnarray}
 \mathbf{H}_{\mathrm{T}}=\sum_{\mathbf{k},\eta \in \mathrm{S,D}}
 ( V_{ \eta \mathbf{k}1 } \mathbf{c}^{\dag}_{\eta \mathbf{k}}
   \mathbf{d}_1 + h.c.
   )+\sum_{\mathbf{k},\mathbf{k}'}(We^{i\phi}\mathbf{c}^{\dag}_{\mathrm{D}\mathbf{k}'}\mathbf{c}_{\mathrm{S}\mathbf{k}}+h.c.),\nonumber\\
\end{eqnarray}
where the first part is the same as that in Eq. (\ref{Equation:H})
while the second part depicts a direct tunneling between the source
and drain leads. $W$ is the lead-lead transmission amplitude assumed
to be independent of momentum $\mathbf{k}$ and $\mathbf{k'}$. $\phi$
is the flux threading the AB-ring, which can be assumed to be
affiliated only to the lead-lead coupling due to the gauge
invariance.

The current through an AB-ring with a dot embedded can be expressed
as\cite{Hofstetter2001, Ueda2006}
\begin{equation}\label{abringcurrent}
    I=\frac{2e}{h}\int d\omega [f_{\mathrm{S}}(\omega)-f_{\mathrm{D}}(\omega)]T(\omega),
\end{equation}
where the transmission probability
\begin{eqnarray}\label{transmission}
 T(\omega)=\frac{4\xi}{(1+\xi)^2}+\frac{2\mathrm{Re}G^{r}_d(\omega)}{(1+\xi)^3}
   [2(1-\xi)\sqrt{\xi\Gamma_{\mathrm{S}}\Gamma_{\mathrm{D}}}\cos{\phi}]\nonumber\\
   +\frac{2\mathrm{Im}G^{r}_d(\omega)}{\Gamma(1+\xi)^3}[(\xi(\Gamma_{\mathrm{S}}^2+\Gamma_{\mathrm{D}}^2)-\Gamma_{\mathrm{S}}\Gamma_{\mathrm{D}}(1+\xi^2)
+4\xi\Gamma_{\mathrm{S}}\Gamma_{\mathrm{D}}\cos^2\phi].\nonumber
\end{eqnarray}
Here $\Gamma_{\eta\in \{\mathrm{S,D}\}}=2\pi
\sum_{\mathbf{k}}|V_{\eta \mathbf{k}1}|^2\delta(\varepsilon_{\eta
\mathbf{k}}-\omega)$ denotes the coupling between the dot and the
source (drain) lead,
$\Gamma=\Gamma_{\mathrm{S}}+\Gamma_{\mathrm{D}}$. The dimensionless
quantity $\xi=\pi^2\nu_{\mathrm{S}}\nu_{\mathrm{D}} W^2$ defines the
direct tunneling coupling between two leads, and
$\nu_{\mathrm{S/D}}$ stands for the density of states of the source
or the drain lead, respectively. $\mathrm{Re}G^r_d(\omega)$ and
$\mathrm{Im}G^r_d(\omega)$ are the real and imaginary parts of the
dot retarded Green function defined as
\begin{eqnarray}
    G^r_d(\omega)=-i\int dt e^{i\omega t}\theta(t)\langle
    \{d_1,e^{iHt}d^{\dag}_1e^{-iHt}\}\rangle.
\end{eqnarray}
Employing the same treatment as to the double-dot interferometer, we
finally obtain the imaginary part of the dot retarded Green function
in the presence of EPI as
\begin{eqnarray}\label{i[lessbar-greatbar]}
   &&
   \mathrm{Im}G^r_{d}(\omega)=-\frac{i}{2}[G^>_d(\omega)-G^<_d(\omega)]\nonumber\\
   &=&-\frac{i}{2}\sum_n L_n [\overline{G}^>_d(\omega-n\omega_0)-\overline{G}^<_d(\omega+n\omega_0)],
\end{eqnarray}
where the decoupled lesser and greater Green functions are
\begin{eqnarray}\label{lesser_bar}
    \overline{G}^<_d(\omega)
=\frac{i|\overline{G}^r_d(\omega)|^2}{(1+\xi)^2} \left[2
\sqrt{\xi\overline{\Gamma}_{\mathrm{S}}\overline{\Gamma}_{\mathrm{D}}}\sin{\phi}
    (f_{\mathrm{S}}-f_{\mathrm{D}})\right.\nonumber\\
  \left.
 +(\overline{\Gamma}_{\mathrm{S}}+\xi\overline{\Gamma}_{\mathrm{D}})
    f_S
    +(\overline{\Gamma}_{\mathrm{D}}+\xi\overline{\Gamma}_{\mathrm{S}})
    f_{\mathrm{D}}
    \right],
\end{eqnarray}
and
\begin{eqnarray}\label{great_bar}
    \overline{G}^>_d(\omega)
    =-\frac{i|\overline{G}^r_d(\omega)|^2}{(1+\xi)^2}
    \left[- 2 \sqrt{\xi\overline{\Gamma}_{\mathrm{S}}\overline{\Gamma}_{\mathrm{D}}}\sin{\phi}
    (f_{\mathrm{S}}-f_{\mathrm{D}})\right.\nonumber\\
    \left.+(\overline{\Gamma}_{\mathrm{S}}+\xi\overline{\Gamma}_{\mathrm{D}})(1- f_{\mathrm{S}})(\overline{\Gamma}_{\mathrm{D}}+\xi\overline{\Gamma}_{\mathrm{S}})(1-
    f_{\mathrm{D}})\right].
\end{eqnarray}
Here the module square of the decoupled retarded Green function is
\begin{equation}\label{ReGrbar}
    |\overline{G}^r_d(\omega)|^2=\left\{[\omega-(\epsilon_0-\frac{\lambda^2}{\omega_0})
    +\frac{\sqrt{\xi\overline{\Gamma}_{\mathrm{S}}\overline{\Gamma}_{\mathrm{D}}}}{1+\xi}\cos\phi]^2
    +(\frac{1}{2}\frac{\overline{\Gamma}_{\mathrm{S}}+\overline{\Gamma}_{\mathrm{D}}}{1+\xi})^2\right\}^{-1},\nonumber
\end{equation}
and the renormalized dot-lead coupling reads
\begin{equation}
    \overline{\Gamma}_{\mathrm{S(D)}}=\Gamma_{\mathrm{S(D)}}\mathrm{exp}[-2(\lambda/\omega_0)^2(N_{\mathrm{ph}}+1/2)].
\end{equation}
Then, by the Kramers-Kronig relation
\begin{equation}\label{ImG to Re}
   \mathrm{ Re}
   G^r_d(\omega)=-\int\frac{d\omega'}{\pi}\mathrm{Re}[\frac{\mathrm{Im}G^r_d(\omega')}{\omega-\omega'+i0^+}].
\end{equation}  The real part of the retarded Green function in
Eq.~(\ref{abringcurrent}) can be easily obtained.


\end{document}